\documentclass{PoS}
\usepackage{graphicx}
\usepackage{subfigure}
\usepackage{gensymb}
\usepackage{siunitx}
\usepackage{comment}

\title{Performance of the INFN Camera calibration device of the first Large Size Telescope in the Cherenkov Telescope Array}

\ShortTitle{Commissioning INFN CaliBox for the first LST camera in CTA observatory}

\author{{M. Palatiello}\thanks{michele.palatiello@ts.infn.it}\,\,$^1$, M. Iori\,$^2$, F. Cassol\,$^3$, D. Cauz\,$^1$, F. Ferrarotto\,$^2$ \\
        $^1$ INFN and University of Udine, Italy \\
        $^2$ INFN Sezione di Roma1 and Rome Sapienza, Italy\\
        $^3$ Aix Marseille Univ, CNRS/IN2P3, CPPM, Marseille, France}
\author{for the CTA Consortium\footnote{https://www.cta-observatory.org/about/cta-consortium/}}

\abstract{On October $10^{\rm th}$ 2018 started the commissioning of the first Large Size Telescope (LST) prototype at the Cherenkov Telescope Array (CTA) northern site at the Observatorio del Roque de los Muchachos, Canary Island of La Palma (Spain). For a precise event energy reconstruction, an LST camera requires a uniform and constant calibration over a large dynamic range, up to $10^{4}$ photo-electrons (p.e.), for each camera photomultiplier tube (PMT). This paper describes the performance of the LST-1 camera calibration system (named CaliBox) in the first commissioning period and provides preliminary results of measurements of the light flat field.}

\FullConference{36th International Cosmic Ray Conference -ICRC2019-\\
		July 24th - August 1st, 2019\\
		Madison, WI, U.S.A.}

\begin{document}

\section{Introduction}
A regular calibration of the LST camera is crucial for an accurate reconstruction of a single Cherenkov event. The CaliBox developed for this purpose is mainly equipped with a 355\,nm UV pulsed laser and two filter wheels to guarantee an adequate photon dynamic range for each PMT over the camera plane 28 m away \cite{doc2,doc1}. 
The light pulses are \SI{400}{\pico\second} wide (FWHM) with an adjustable rate between \SI{1}{\hertz} and \SI{2000}{\hertz}. The CaliBox is equipped with a set of filters to achieve at least the required dynamic range from 10 to $10^4$ p.e. at the PMTs as recommended to ensure a long PMT lifetime. At every shot the laser provides a trigger to the camera, while in addition an external trigger can be generated by the ODROID-C1+ General Purpose Input/Output (GPIO) board \cite{odroid}. All the system is completely remotely managed by a properly designed Media Object Server Communications Protocol (MOS) \cite{mos} based on an Open Platform Communication-Unified Architecture (OPC-UA) Server running on an ODROID-C1+ board, as the central managing system of the CaliBox devices and sensors.

\section{Description and performances of CaliBox}
The CaliBox structure consists mainly of two optically connected, nitrogen filled and hermetically closed (IP67 certified) aluminum boxes fixed on a large aluminum plate to properly dissipate the heat generated by all the internal components. The larger aluminum box contains the laser and the filter wheels while the smaller one contains a 1 inch integrating sphere and two different photon sensors. Outside the two boxes and fixed on the aluminium plate, are placed a \SI{100}{\watt} power supply and the ODROID-C1+ computer board. All this internal structure is contained in an aluminum water proof shell (see Fig.\,1a).
\begin{figure}[htbp]
        \centering%
        \subfigure[]%
          {\includegraphics[width=65mm]{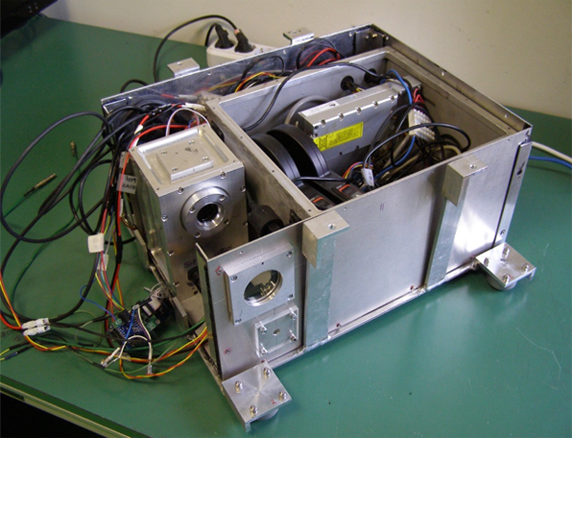}}\qquad
        \subfigure[]%
          {\includegraphics[width=70mm]{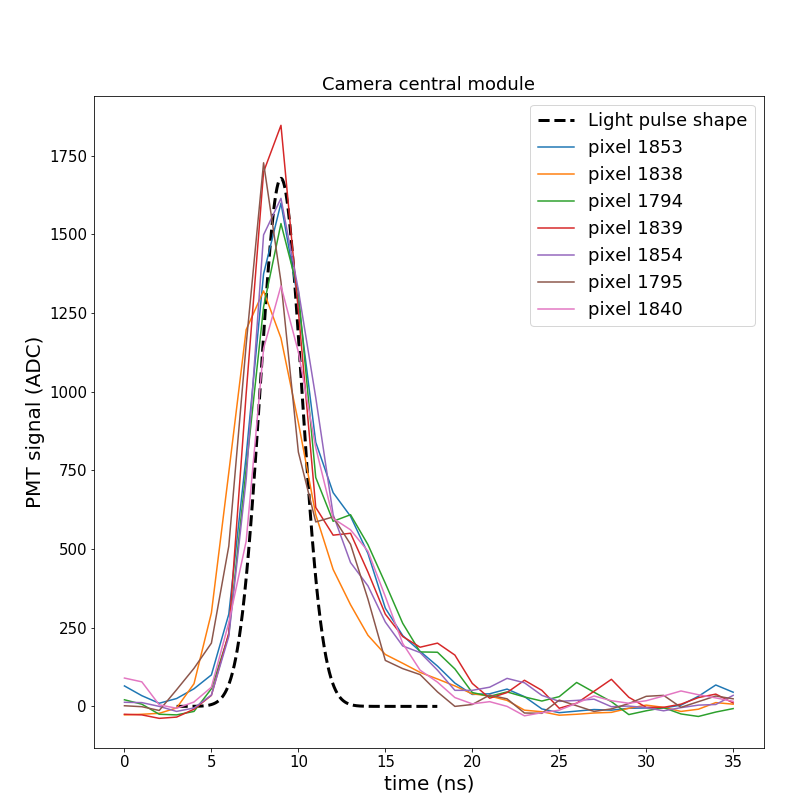}}
        \caption{(a) CaliBox internal boxes and external shell. In this picture the CaliBox is inserted in the support for a secure and easy installation at the center of the telescope mirror dish. (b) Black dot line: pulse shape at the exit of the integrating sphere (FWHM of $2.80\pm0.03\,$\si{\nano\second}). Color lines: a few LST Camera PMT signals.}
\end{figure}
The presence of two different photon sensors (Photodiode Hamamatsu 3590-18 (PD) and a Sensl 3x3 \si{\square\milli\metre} Silicon-Photomultiplier (SiPM)), at one exit port of the diffuser, ensures the monitoring of photon intensity over all the requested dynamic range, the SiPM for low and the PD for high photon intensity, through the use of a 10 bit ADC. That allows an inter-calibration of the number of photons sent to the camera. 
Measurement of the signal shape at the exit of the sphere results in a FWHM of $2.80\pm0.03\,$\si{\nano\second}, consistent with the Cherenkov signal shape produced by air showers in a single PMT \cite{doc2}. 
The CaliBox light uniformity was tested using a SiPM successively located at 27 different points on a plane 5\,m away form the device covering the entire camera PMT's solid angle and a DRS4 \cite{drs4} to read and integrate the signal to an accuracy of 2\% in the light intensity (see Fig.\,2a). 
The CaliBox was installed in December 2018 and since then we have been monitoring the internal and external relative humidity and temperature during the data taking in the first semester of the year. The CaliBox results to be a very stable device from a mechanical, performance and difficult environmental conditions (dust, ice, strong wind) point of view.
\begin{figure}[htbp]
        \centering%
        \subfigure[]%
          {\includegraphics[width=73mm]{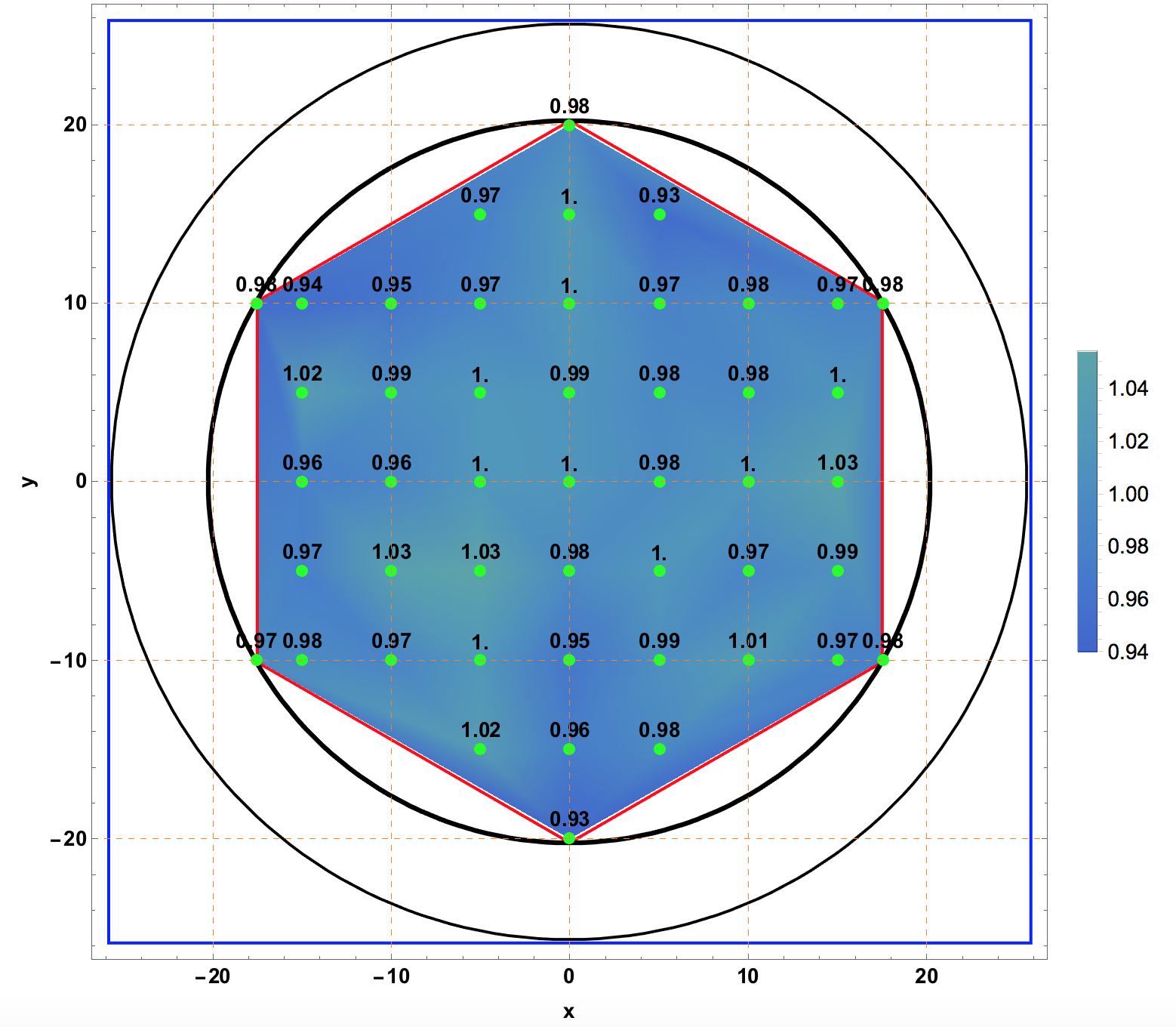}}\qquad
        \subfigure[]%
          {\includegraphics[width=70mm]{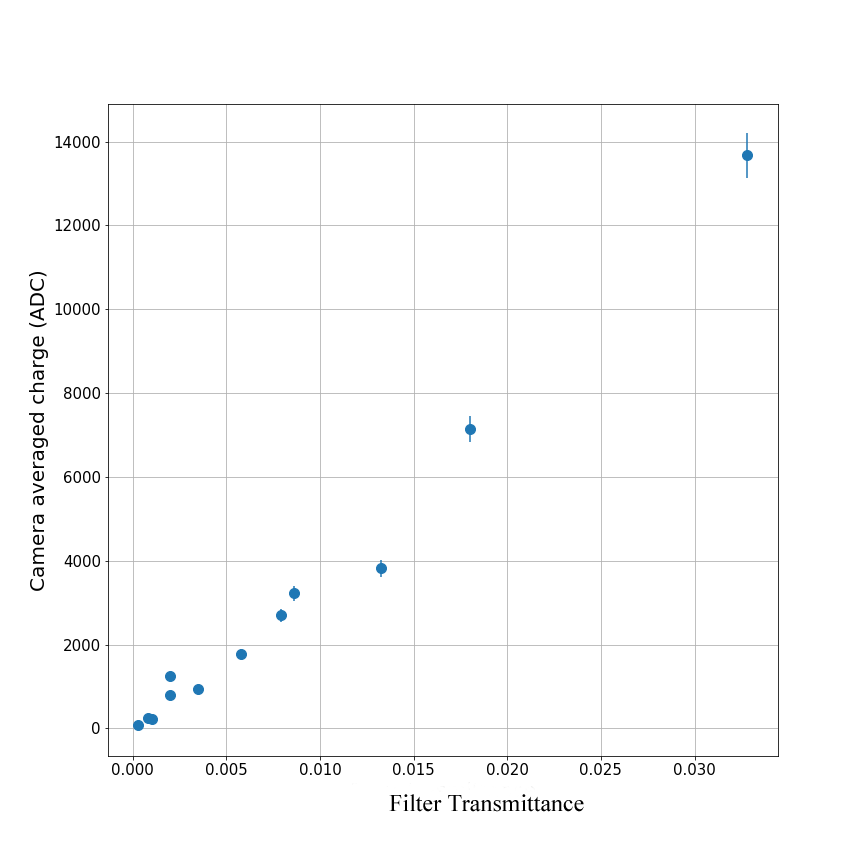}}
        \caption{(a) CaliBox uniformity study results performed in laboratory using a screen located 5\,m away from the device. A SiPM was successively placed at 57 different points of coordinate x,y in centimeters, to cover the entire camera PMT's solid angle. For each position the intensity relative to the central point value is given, showing a variation around  2\%. The hexagonal shape of the LST camera is shown in red. (b) Averaged charge of the camera PMTs as a function of filter transmittance obtained by different filter combinations.}
\end{figure}

\section{The OPC-UA Server protocol}
The CaliBox is one of the auxiliary systems of the LST Camera that has to be directly managed by the Central Camera Control software by an OPC-UA Client.
Inside this scenario all the CaliBox managing is operated by a single-board ODROID-C1+ computer over a Ubuntu Linux operating system. Every device inside the box can be run and debugged directly by executable files (C++ compiled) and at the same time can be managed, via Secure SHell protocol, by the MOS based on OPC-UA Server in calling state inside the ODROID-C1+ board. 
The MOS protocol allows to convert all the C++ programs in plugins that are completely managed by the OPC-UA Sever for a direct access via Methods and DataPoints by an OPC-UA Client \cite{client} independently of external system. From the OPC-UA Client, it is possible, e.g., to set the laser parameters (frequency, duty cycle and number of shots), change the filter combination, start and stop the red pointing laser and get the value of temperature and relative humidity inside the box. It is also possible to generate a periodic, random or Poisson-distributed pedestal trigger completely managed (type, frequency, duty cycle and number of shots) as a Method or DataPoint by the OPC-UA Client.

\section{Conclusions}
During the first six months of data taking under severe weather conditions (strong wind, ice, dust), the CaliBox has shown a correct and stable behavior from the mechanical, optical and performance point of view.
Within the CTA scenario of a new generation of gamma-ray observatory, our system fulfills all requests for a continuous calibration of the LST camera uniformity,  by providing an appropriate pulse shape and photon densities at the camera plane corresponding to a large (10 to $10^{4}$ p.e.) dynamic range.
The HV flat field camera PMT has been obtained within 2\% using the light flat field produced by the laser on the camera plane. The acquisition of interleaved flat-field events ($\sim$100 Hz) will be soon employed for the absolute calibration of the camera on the base of the photon-statistic (F-Factor) method \cite{Franca}.
The ODROID-C1+ is equipped with a calling OPC-UA Server protocol with a dedicated addition to generate a pedestal trigger as well.
\\
\section*{\centering ACKNOWLEDGMENTS}
\small {We are grateful to P. Gardonio and A. Kras from the DPIA Dept. of the University of Udine, for the CaliBox mechanical vibrations study, and for the CAD drawings. We acknowledge as well the assistance of S. Saggini (DPIA) for some suggestions on the Trigger Interface Board (TIB) circuit. We would also like to thank Dr. L. Recchia, G. Chiodi, R. Lunadei of INFN-RM1 Electronics Laboratory (LABE) and A. Sabatini of University and INFN of Udine Mechanical Workshop.
We gratefully acknowledge financial support from the agencies and organizations listed here: http://www.cta-observatory.org/consortium$\_$acknowledgments. This work was conducted in the context of the CTA LST project.}


\begin{thebibliography}{99}
\bibitem{doc2}
Technical Design Report for LST, see https://www.cta-observatory.org/
\bibitem{doc1}
M. Iori et al. for the CTA Consortium, PoS ICRC2015 (2016) 954
\bibitem{odroid}
https://www.odroid.co.uk/hardkernel-odroid-c1
 \bibitem{mos}
Media Object Server Communications Protocol (MOS) provided by the Laboratoire d'Annecy-le-Vieux de Physique des Particules (LAPP)
\bibitem{drs4}
https://www.psi.ch/drs/DocumentationEN/manual$\_$rev50.pdf
\bibitem{client}
https://www.unified-automation.com/products/development-tools/uaexpert.html
\bibitem{Franca}
Bencheikh et al 1992, NIMA 315 (13), 349-353


\end{thebibliography}
\end{document}